\documentclass[aps,nofootinbib,showpacs,preprint]{revtex4}
\usepackage{xcolor}
\usepackage[hypertex]{hyperref}
\usepackage{graphicx,subfigure}
\usepackage{amsmath,amsfonts,amssymb}

\newcommand{\f}{\frac}
\newcommand{\suml}{\sum\limits}

\newcommand{\eql}[1]{Eq.~\eqref{#1}}
 \newcommand*{\OrigAA}{}
\let\OrigAA\AA

\renewcommand*{\AA}{%
  {\fontfamily{ptm}%
  \selectfont%
  \OrigAA%
  \selectfont}%
}
\begin{document}
\title{Is the thermodynamic behavior of the noble fluids consistent with the Principle of Corresponding States?}
\author{V.L. Kulinskii}
\altaffiliation[On leave from the ]{Department of Theoretical Physics, Odessa National
University, Dvoryanskaya 2, 65026 Odessa, Ukraine}
\affiliation{Department of Molecular Physics, Kiev National
University, Glushko 6, 65026 Kiev, Ukraine}
\email{kulinskij@onu.edu.ua}
\author{N.P. Malomuzh}
\email{mnp@normaplus.com}
\author{O.I. Matvejchuk}
\affiliation{Department of Theoretical Physics, Odessa National
University, Dvoryanskaya 2, 65026 Odessa, Ukraine}
\begin{abstract}
The applicability of the Principle of Corresponding States (PCS)
for the noble fluids is discussed. We give the thermodynamic
evidences for the dimerization of the
liquid phase in heavy noble gases like argon, krypton etc.
which manifest itself in deviation from the PCS.
The behavior of the rectilinear diameter of the entropy
and the density is analyzed. It is shown that these
characteristics are very sensitive to the dimerization
process which takes place in the liquid phase of heavy noble gases.
\end{abstract}
 \maketitle
%\linenumbers
\section{Introduction}
The principle of thermodynamic similarity is one of the fundamental
guiding principle in investigation the thermodynamic properties of
matter. It allows to determine the classes of the substances with
similar thermodynamic behavior based on the similarity of the
interparticle interactions, which in its turn is based on the
similarity of the particles (electronic shells, molecular excitation
spectra etc.) and as a sequence the similarity of the interparticle interactions.
The existence of the universality classes of critical
behavior can be considered as the extension of the thermodynamic
similarity principle to the long range fluctuations
%rather than the particles
since the short range effects does not change the
asymptotic behavior of the thermodynamic quantities.
In particular, for the systems obeying the the thermodynamic similarity principle the
dimensionless critical amplitudes should coincide.
The demand of the thermodynamic similarity and
as the consequence the principle
of corresponding states (PCS) is very strong condition which in fact
assumes the similarity of the energy spectrum of the systems since the
thermodynamics depends on both the external and the internal
molecular degrees of freedom.

It is well known that the thermodynamics is determined by the
integral characteristics of the interaction potential. The
classical PCS states the trivial fact that the systems with  if
the interaction potential is of two-particle character and has the
form:
\begin{equation}\label{potential}
U = U_0 f(r/r_0)\,,
\end{equation}
where $U_0$ is the depth of the potential well and $r_0$ is the characteristic molecular size correspondingly.
Then the corresponding equation of state (EOS) has the universal dimensionless form
\[p^* = p\left(\,v^*,T^*\,\right)\,,\]
where $p^* = Pr^3_0/U_0$ is the dimensionless pressure $P$,
$T^* = T/U_0$ and $v^* = v/r_0^3$ are the dimensionless
temperature $T$ and the specific volume $v$ correspondingly.
The famous van der Waals (vdW) equation of state is the illustration of the PCS.
Throughout the paper we use the units where $k_B = 1$.
In particular the locus of the critical point (CP) depends on the
potential depth $U_0$ and the molecular radius (hard core radius)
%.In particular for the coexistence curve $p(t)$
and it is expected that the locus of the triple points are
the same too.

Sure that such a simple form of the interparticle interaction is
nothing than oversimplification since the classical potential $U(r)$
is the effective interaction. It is obtained via integrating out the
electronic degrees of freedom and in fact becomes dependent on the
atomic number $A$ and the de Boer parameter $\Lambda
=\f{h}{r_0\sqrt{m U_0}}$ (note that $m = A \,m_p$, where $m_p$ is
the proton mass). For the noble fluids the latter varies in
wide range from $\Lambda_{He} = 2.7$ to $\Lambda _{Xe} = 0.06$
\cite{book_prigozhisolut,book_hansenmcdonald,dimers_hirsch/jcp/1959}.

Nevertheless the quantum effects manifest themselves in dual way:
a) the corrections to the asymptotic of classical interaction
\cite{book_kaplan}; b) the formation of bound states (e.g. dimers,
trimers etc.). In the first case the potential conserves its
structure in essential. The bound states lead to the additional
contributions to the basic  thermodynamic quantities (the specific
volume, heat capacity etc.). The system becomes the mixture of
monomers, dimers etc. The formation of dimers in noble gases is known
\cite{dimers_ogilvie/jmolstruct/1992}. They manifest itself in various effects based on the anisotropy structure of the dimers, e.g. in Raman spectra \cite{dimers_argon_raman/prl/1972,dimers_argon_raman/jcp/1975,dimers_raman_Godfried/prl/1982}, depolarization of the light scattering \cite{trojanovsky_phd,dimers_neon_depolarization/pra/1994} etc. But their existence in liquid phase near the CP has not been discussed. Note that the dimers have finite lifetimes but in equilibrium there is certain amount of them.

As it follows from the Table~\ref{tab1} the number of the vibrational levels $N_{vib}(U_0)$ is different for heavy noble gases. The dimerization breaks the PCS. The manifestation of this effect is expected in gaseous phase of low density (see below). But the degree of dimerization is low for dense liquid states. Only in the vicinity of the critical point the contribution from the dimers could be great enough to break the applicability of the PCS.

%This allow to use the scaling transformation of the potentials \eqref{potential}. If the potentials differs very much by the parameter $U_0$ then the homogeneity property based on Eq.~\eqref{quasiclass} is broken and the difference in the dimers ratio lead to the declination of the PCS for such substances even though the potentials are of the same form, e.g. Lenard-Jones type. In fact this is well known fact that heavy noble fluids like $Ar,Kr,Xe$ falls in the same class of the corresponding states, while the lighter ones $He,H_2,Ne$ are dimerized to a negligible degree and differ in this respect from heavy noble fluids.
%The results for the characteristics of the dimers of different noble fluids are summarized in Table~\ref{tab1}. It is clear that in case of $\left(\,H_2\,\right)_2$ and $Ne_2$ the inequality \eqref{quasiclass} becomes invalid.
%
\begin{table}
  \centering
\begin{tabular}{|c|c|c|c|c|c|c|}
\hline
   & $He$ & $H_2$ & $Ne$ & $Ar$ & $Kr$ & $Xe$ \\
   \hline
% $r_a, \AA$& 0.31& & 0.38& 0.71& 0.88& 1.08\\ \hline
   $r^{vdW},\text{\AA} $& 1.40 & 1.54 &1.54& 1.88& 2.02& 2.16\\ \hline
     $l_{bond},\text{\AA} $ \cite{book_kaplan,dimers_zhao/jphyschem/2006} & - & 4.2 &3.10 &3.76 &4.01 & 4.36 \\ \hline
$\varepsilon_{diss}/T_c$, \cite{dimers_ogilvie/jmolstruct/1992,dimers_zhao/jphyschem/2006}& 0.0 & 0.11 & 0.95 & 0.96 & 0.96 & 0.97\\ \hline
%$v_{c}/v_{vdW}$ & 8.3& 7.0& 2.5& 3.6& 3.6& 2.1\\ \hline
$Z_{c}=\f{P_c v_c}{T_c}$ & 0.30& 0.30& 0.30& 0.29& 0.29& 0.29\\ \hline
$N_{vib}$, \cite{dimers_hirsch/jcp/1959} & - & 2 & 3& 8& 13& 21\\ \hline
\end{tabular}
  \caption{Basic geometric and energetic parameters of the dimers and the molecules. Notations: $r^{vdW}$ is the van der Waals radius of the molecule, $l_{bond}$ bond lentght of the dimer, $\varepsilon_{diss}/T_c$ is the dissociation energy of the dimer in units of $T_c$, $Z_c$ is the compressibility factor.}\label{tab1}
\end{table}
Therefore the equation of state (EOS) takes the form:
\begin{equation}\label{press}
p^* = p(T^*,v^*_{1},v^*_{2},\ldots) +
p_{q}(T^*,\Lambda,v^*_{1},v^*_{2},\ldots)\,,
\end{equation}
where subscript ``q`` stands for quantum correction and
$v^*_i$ are the dimensionless specific volumes of monomers, dimers, etc.

Because of the complete electronic shells and as a sequence the
spherical symmetry of the ground state the noble fluids are ideal
substances for the test of the PCS.

The aim of this paper is the analysis of the applicability of the
PCS for the noble fluids and the interpretation of the data from
the point view of dimerization processes in these substances. More
exactly we analyze 1) the applicability of the PCS for the
description of the noble gases on their coexistence curves; 2) the
manifestation of the dimerization effects on the temperature
behavior of their the rectilinear diameters; 3) the influence of
the dimerization on the peculiarities of the critical
fluctuations.

In accordance with this Section~\ref{sec_exp_facts} is devoted to
the careful analysis of the experimental data (all the data are from
NIST open database). In Section~\ref{sec_ref_sys} the purposeful
choice of the reference system is discussed. In
Section~\ref{sec_dimer} we consider simple model to account the
dimerization process for noble fluids. The mean field behavior of
the rectilinear diameters are considered in
Section~\ref{sec_diam1} and the critical anomaly is discussed in
Section~\ref{sec_diam2}.

\section{Experimental facts}\label{sec_exp_facts}
In this Section we discuss: 1) the comparable behavior of the
coexistence curves (CC) for noble fluids ; 2) the behavior of the
number density along the CC.
Further we will use the variables reduced to the CP values:
\[\tilde{p} = P/P_c,\, \quad \tilde{T} = T/T_c,\,\quad \tilde{n} = n/n_c\,.\]

\begin{figure}
  \centering
  \includegraphics[scale=0.7]{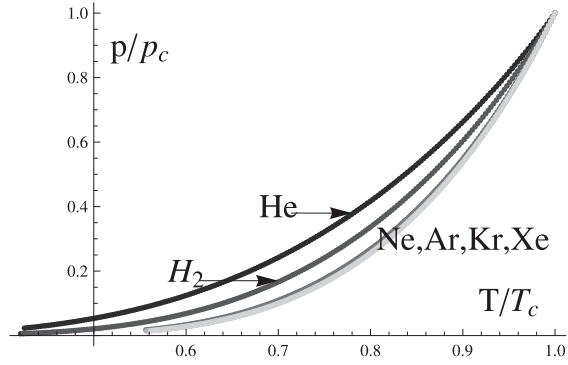}\\
  \caption{The CC in $\tilde{p}-\tilde{T}$ coordinates.}\label{pt}
\end{figure}

First we consider the role of the quantum effects in the EOS. On
Fig.~\ref{pt} we give the CC in ``$\tilde{p}-\tilde{T}$``
coordinates for classical the noble fluids ($Ne,Ar,Kr,Xe$) and the
quantum ones ($He$ and $H_2$).
We see that in ``$\tilde{p}-\tilde{T}$`` coordinates the classical
noble fluids obey the PCS. The deviation from the classical
behavior is connected with the quantum effects in accordance with
Eq.~\eqref{press}. They are greater for $He$ than for molecular
hydrogen. Note that ``$\tilde{p}-\tilde{T}$`` diagram is the
robust thermodynamic characteristics which does not allow to
distinguish fine details of the thermodynamic behavior. The
variables $\tilde{n},\tilde{T}$ are more suitable in this sense.

\begin{figure}
    \centering
  \includegraphics[scale=0.75]{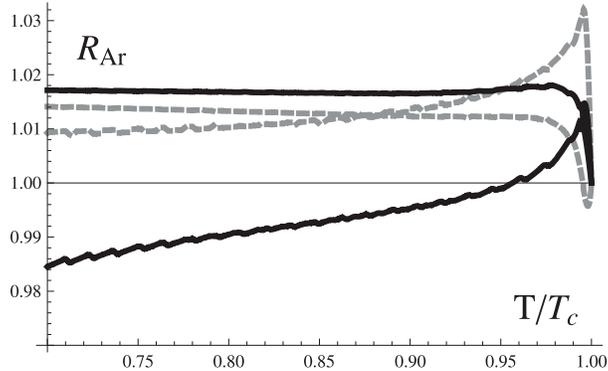}\\
  \caption{The ratio $R(Kr,Ar)$ (dashed) and $R(Xe,Ar)$ (bold)}\label{densrefar}
\end{figure}

Let us consider the dimensionless ratio $R(2,1) = \tilde{n}_2/\tilde{n}_1$ of
the dimensionless densities of noble fluid ''2'' to the density of
noble fluid ''1'', which in some reason is chosen as the reference
fluid. The results of comparison for the liquid and gaseous
branches of $Kr$ and $Xe$ with respect to $Ar$ are presented in
Fig.~\ref{densrefar}. As is seen from it the deviation from unity for liquid
branch within the range of the experimental errors ($<1\%$). For
the gaseous phase the difference  between the $Kr$ and $Xe$ curves
reaches 3\%.

The deviation from unity can be considered as the measure of the
violation of the PCS. In accordance with said above the deviation
is noticeable in the gaseous phase where the dimerization is more
explicit. The physical ground for the existence of the dimers in
near critical region is the almost free rotation of the dimers
allowed at low enough density. As is known this  leads to the
existence of the metastable states for the dimers with a wide
range of life time because of the centrifugal barrier
\cite{dimers_hirsch/jcp/1959,dimers_baylis/pra/1970}.
In liquid phase the deviation becomes essential in the near critical region.
In order to clarify the physical nature of such deviation we take
into account that along the CC the density is the sum of classical
and quantum contributions. According to \eql{press} for the
density we have:
\begin{equation}\label{dens}
  \tilde{n}  = \tilde{n}_{cl}(\tilde{T}) + \tilde{n}_{q}(\tilde{T};\Lambda )
\end{equation}
Note that the quantum correction $n_{q}(\tilde{T};\Lambda )$ is
monotonous decreasing function on $T$. As is follows from the
results presented in Fig.~\ref{pt} for classical noble fluids the
quantum corrections are negligibly small. Therefore the deviation
$R$ from 1 in liquid and gaseous phase is naturally to connect
with the dimerization. In order to estimate the dimerization
degree it is useful to search for the reference fluid in which the
dimerization is depressed.

In the following section we show that
molecular hydrogen or $Ne$ can be chosen that such fluid.
\section{Reference system}\label{sec_ref_sys}
Let us give the arguments that 1) the molecular hydrogen $H_2$ is
the simplest semiquantum fluid and 2) $Ne$ is the simplest
classical fluid. This means that in liquid and gaseous phases of these fluids in near critical region the amount of the dimers is very small and does not affect the thermodynamic behavior. It becomes clear from
the value of the dissociation energy $\varepsilon_{diss}$ reduced
to the critical temperature $T_c$. In molecular hydrogen the
dissociation energy for its dimer is an order of magnitude smaller
than the critical temperature $\epsilon_{diss}/T_c \approx 0.11$. In $He$ no dimers exists except for the giant states \cite{dimers_helium_halo/europhys/2007} which is out of the scope of the paper.
%Helium is specific case since the formation of the giant dimers
%takes place in gaseous and possible in liquid phase too
%\cite{dimers_helium_cahn/prl/2008}.
For $Ne$ the dimerization is
weak in comparison to other classical noble fluids. $Ne$ is the
closest to $H_2$ among the classical noble fluids. The comparison
assumes that the classical constituent of the density for $Ne$ and
$H_2$ is different from that of heavy noble gases. Note that their
critical temperatures is pretty close $T_c \sim 44.5\,K$ and $T_c \sim 33.\,K$ correspondingly and high
enough so that the difference of these light noble gases from the
heavy ones can not be attributed to quantum corrections which
negligible in near critical region.

In accordance to Table~\ref{tab1} the formation of the dimers
takes place mainly in classical noble fluids. Though the formation
of dimers for $Ne$ is possible the number of vibrational levels is
not big and the dimerization can be neglected in comparison with
other classical noble gases.

\begin{figure}
  \includegraphics[scale=0.75]{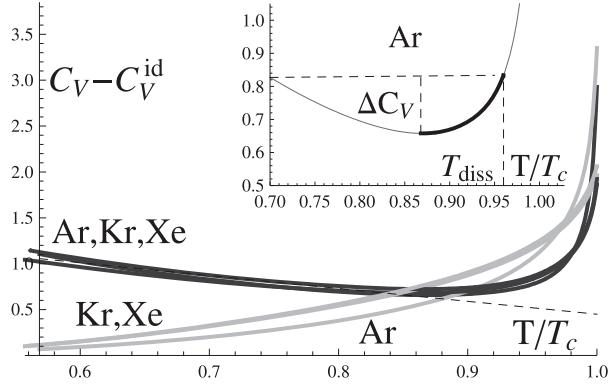}
  %\hspace{1cm}
%  \includegraphics[scale=0.5]{cvargon}
  \\
  \caption{Residual specific heat (in units of $k_B$, $C^{id}_V = 3/2$ is the specific heat of the monoatomic ideal gas) of the noble gases along liquid (black) and vapor (gray) branches of the binodal.
  The inset shows the increment of the specific heat caused by the dimerization in near critical liquid phase (bold segment of the curve).}\label{cv}
\end{figure}

In order to estimate the fraction of dimers in liquid phase we use
the specific (per atom) heat data (see Fig.~\ref{cv}).
Here we see the monotonic change of the residual (after subtraction of the specific heat of the monoatomic ideal gas contribution) specific heat in gaseous phase due to increase of interparticle interaction with the density up to the temperature $T/T_c\approx 0.95$, which has the meaning of the Ginzburg temperature \cite{book_patpokr}. Note that in this phase for $T/T_c \lesssim 0.95$ the specific heat of $Ar$ is noticeably less than that for $Kr$ and $Xe$ due to more strong interpaticle interaction.
The opposite behavior of the specific heat takes place in liquid phase. Since the density of the liquid phase decreases along the binodal the corresponding decrease of the residual specific heat is observed in the temperature region $T/T_c \lesssim 0.95$. For $T/T_c > 0.95$ the specific heat of both phases diverges due to critical fluctuations.

\begin{figure}
\centering
\includegraphics[scale=0.75]{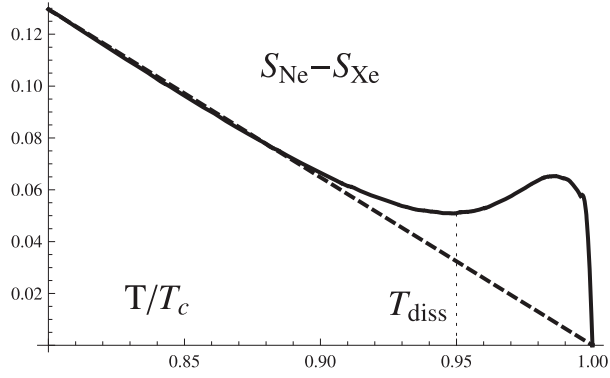}\label{fig_sintarne}
 \caption{The difference of the entropies of the liquid phases of $Ne$ and $Xe$.}\label{fig_liqentrop}
\end{figure}

From the physical point the decrease of the specific heat of the liquid phase outside of the fluctuational region is connected with the decrease of the average interparticle interaction as the density decreases. The existence of the minimum at $T_{min} \approx 0.87$ and the subsequent increase of $C_V$  until $T_{diss} \approx 0.95$ (see Fig.~\ref{cv}) can be attributed to the formation of the dimers.
%%%%%%%%%%%%%%%%%%%%%%%%%%%%%%%%%%%%%%%%%%%%%%%%%%%%%%%%%%%%%%%%%%%%%%%%%%%%%%%
%%%%%%%%%%%%%%%%%%%%%%%%%%%%%%%%%%%%%%%%%%%%%%%%%%%%%%%%%%%%%%%%%%%%%%%%%%%%%%%
Representing the specific heat as
\begin{equation}\label{cvid12}
  C_V = \f{3}{2}\,N_1 + \f{7}{2}\,N_2
\end{equation}
where $N_1$ is the number of atoms and $N_2$ is the number of the dimers,
we get the estimation for the degree of dimerization (the ratio of the number of dimers to the total number of atoms):
\[\f{N_2}{N_0} = \f{2}{7}\f{\Delta C_V}{N_0}\approx 0.05\,.\]
where $N_0 = N_1+2N_2$ is the total number of atoms, $\Delta C_V  = C_V(T_{diss}) - C_V(T_{min})$ is the increment of the specific heat due to the formation of the dimers. Here it is assumed that the dimer has 6 degrees of freedom (3 translational, 2 rotational and one vibrational). Of course the dimerization in liquid phase should manifest itself in the behavior of the entropy too (see Fig.~\ref{fig_liqentrop}). Here we see that up to the temperature $T\approx 0.87$ the monotonous behavior of $S_{Ne}-S_{Xe}$ takes place. It signifies about the absence of the clusterization ordering in $Xe$. Noticeable diminishing the monotonic decrease in the temperature interval $0.87 <T/T_c\le 0.95$ is naturally explained by the dimerization of $Xe$ leading to decrease of its entropy in comparison with that of $Ne$.
%%%%%%%%%%%%%%%%%%%%%%%%%%%%%%%%%%%%%%%%%%%%%%%%%%%%%%%%%%%%%%%%%%%%%%%%%%%%%%
%%%%%%%%%%%%%%%%%%%%%%%%%%%%%%%%%%%%%%%%%%%%%%%%%%%%%%%%%%%%%%%%%%%%%%%%%%%%%%

\begin{figure}[t]
    \centering
  \includegraphics[scale=0.7]{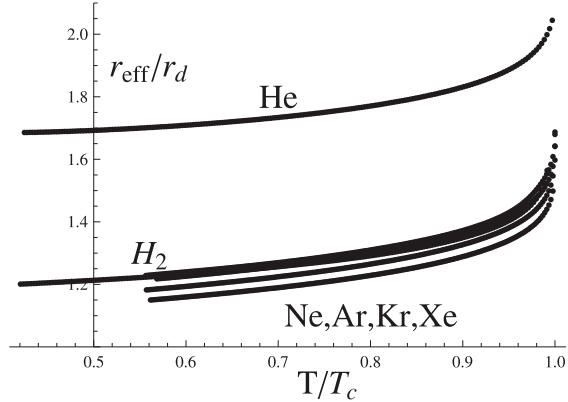}\\
  \caption{The dimensionless distance $\tilde{r} =
  2 \left(\,\f{3v}{2\pi}\,\right)^{1/3}/(l_{bond}+\sigma)$}\label{dist}
\end{figure}

This energetic condition should complemented with the geometric
condition of free rotation of the dimers. Two molecules occupy in
the average the volume $2/n$, therefore he radius of the effective
cell for a dimer is $r_{eff} \simeq \left(\f{3}{2\pi
n}\right)^{1/3}$. The rotating dimer occupies the volume of the
radius $r_{d} = \f{l+\sigma}{2}$. The ratio $r_{eff}/r_{d}$ which
is shown in Fig.~\ref{dist} and becomes greater than unity in the
vicinity of the CP.

\begin{figure}[hbt!]
  \centering
        \subfigure[$He$]{\includegraphics[scale=0.7]{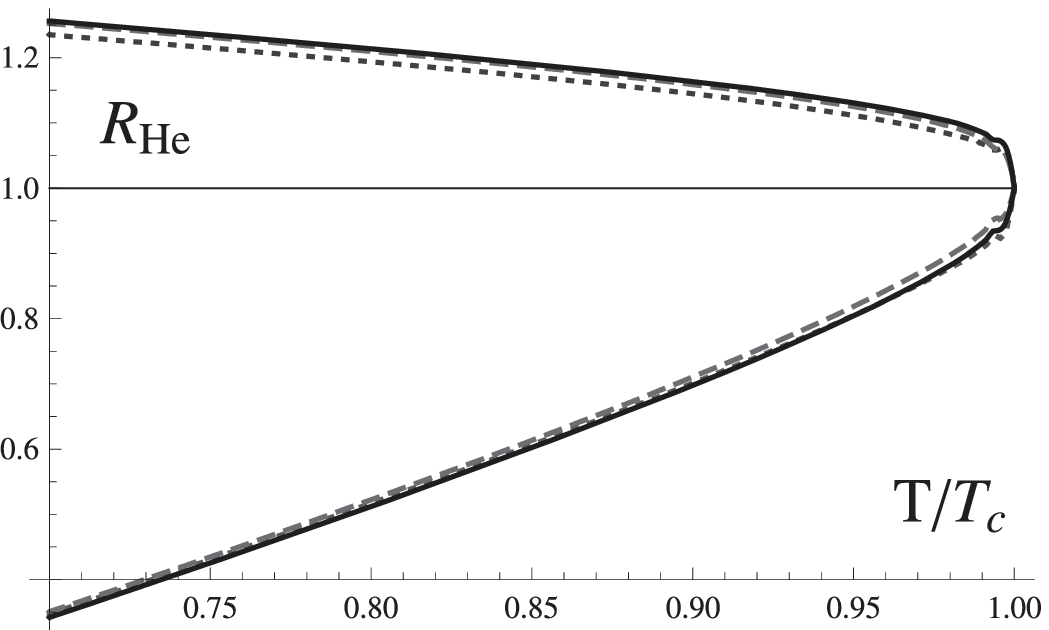}}
  \subfigure[\,$H_2$]{\includegraphics[scale=0.7]{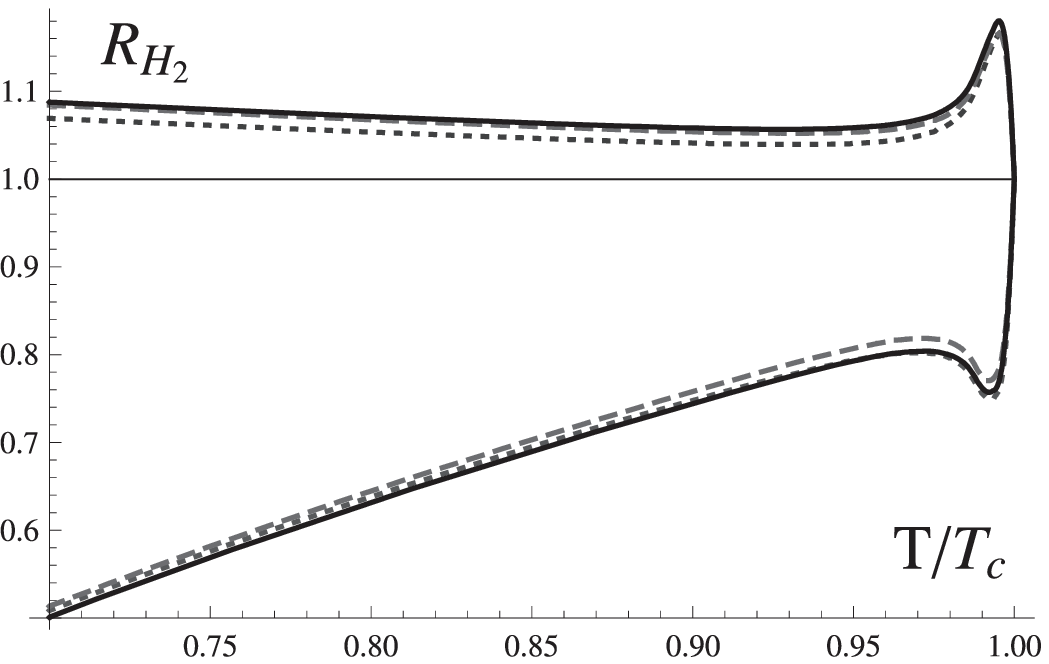}}\,\,
          \subfigure[\,$Ne$]{\includegraphics[scale=0.7]{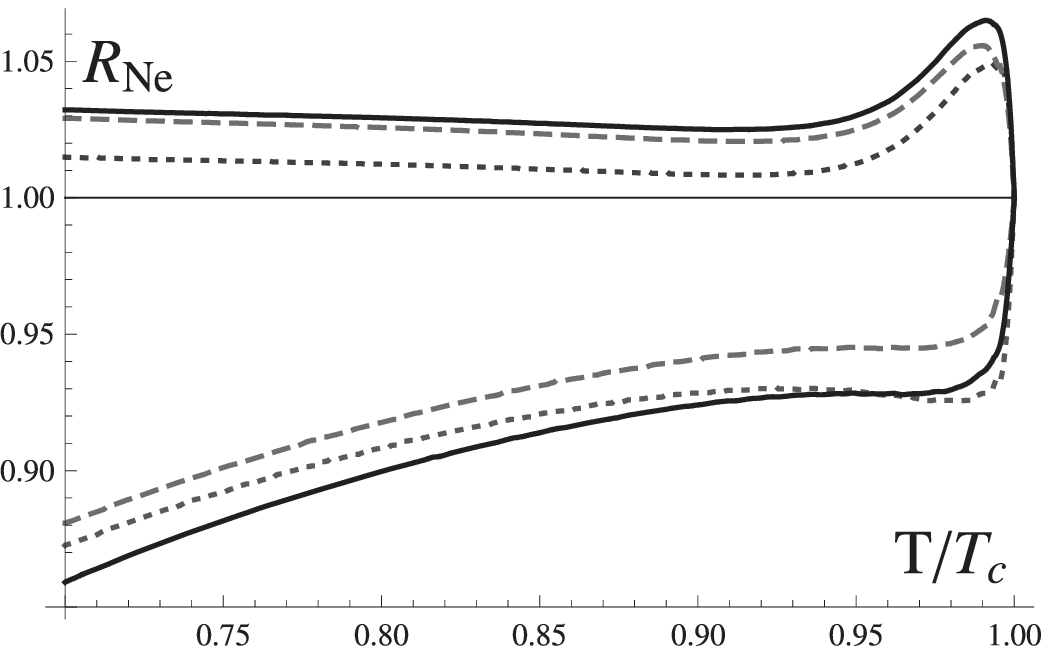}}\,\,
  \caption{The ratio $R_{i}$ along the binodal for $Ar$ (dotted),
  $Kr$ (dashed), $Xe$ (bold) and different reference fluids (see the captions).}\label{nobleref}
\end{figure}

Let us consider one of the light noble fluid as the reference
system. One can conclude that the ratio of the densities for the
heavy and light noble gases:
\begin{equation}\label{ration}
R_{i} = \f{\tilde{n}^{(i)}_{cl}(\tilde{T})}{\tilde{n}^{(ref)}_{cl}(\tilde{T}) + \tilde{n}^{(ref)}_{q}(\tilde{T},\Lambda)}
\end{equation}
is monotonous along the coexistence curve provided that the
classical contributions are the same. In Fig.~\ref{nobleref} we
give the ratio $R$ for heavy noble fluids ($Ar, Kr, Xe$) with
respect to the lighter ones ($H_2, Ne, He$). It shows that in the
near critical region according to the molecular hydrogen is more
like $Ne$ than $He$. The additional support in favor of the
thermodynamic similarity $H_2$ and $Ne$ is the closeness of their
CP locus. Note that in gaseous phase the significant deviation
of $R$ from 1 with lowering the temperature (density of saturated
vapor decreases too) is connected with the dimerization in accordance
with what has been said in the Introduction. From Fig.~\ref{rcompar}
we can conclude that the dimerization degree in gaseous phase is more than $10\%$.

From what has been said above the nonmonotonic change of $R$ can
be attributed to the temperature behavior of the classical part
$\tilde{v}_{cl}$ of the specific volume only. The classical
behavior for the noble fluids is not the same as for $H_2$ because
of formation of the dimers. That mean that the classical part of
the thermodynamic functions corresponds to the monoatomic vdW
fluid.

The dissociation energy for dimers of heavy noble fluids is close
to the dimer bound energy (the critical temperature slightly
($4\%-6 \%$) higher than the dissociation energy of the dimer
$\varepsilon_{diss} /T_c\approx 0.95$, see Table~\ref{tab1})
\cite{dimers_ogilvie/jmolstruct/1992}. From Fig~\ref{nobleref} one
can see that nonmonotonic change in the ratio $R$ occurs right at
the temperature $kT \simeq \varepsilon_{diss}$.

It can be seen that the geometric  (see Fig.~\ref{dist}) and
energetic conditions give the ground to expect the nonzeroth
dimerization of heavy noble fluids in the vicinity of the CP.
One can choose either $Ne$ or molecular hydrogen as the
reference fluid since they can be treated without taking into
account of the dimers $Ne_2$ or $(H_2)_2$. In $Ne$ although the
energetic condition allows the formation of dimers the number of
the bound states is low and its dimerization
in near critical region is negligibly small
in comparison with heavier noble fluids. The spectrum of dimers of
heavy noble fluids have many vibrational states. Thus the dimers
of heavy noble fluid are more stable. From this point of view the
molecular hydrogen and $Ne$ are indeed the ``simplest`` molecular
liquids. The quantum effect for $H_2$ should be taken into account
via the representation Eq.~\eqref{press}.

From here we can conclude that the formation of dimers
in liquid phase takes place
in near critical region where the density is low enough so that
both energetic and geometric conditions are valid. Therefore there
is the difference in short range correlations in light and heavy
noble fluids which is based on the formation of bound clusters.

In fact, the liquid phase of a noble gas is the mixture of
monomers and small quantities of dimers. Since their concentration
depends on the kind of matter, the critical amplitudes for
different noble gases will take values deviating from those, which
follow from the principle of corresponding states.  In this sense
one can say that the quantum properties, namely the stability
properties of bound states, influence the critical parameters of
the classical noble fluids. This result was noticed in numerical
simulations \cite{dimers_quantumfluct/pra/1997}.

%%%%%%%%%%%%%%%%%%%%%%%%%%%%%%%%%%%%%%%%%%%%%%%%%%%%%%%%%%%%%%%%%%%%%%%%%%%%%%%%%%%%%%%%%%%%%%
\section{The behavior of the rectilinear diameter}\label{sec_diam1}
The rectilinear diameter for the density:
\begin{equation}\label{diamd}
  n_{d} = \f{n_{l}+n_{g}}{2n_c} - 1
\end{equation}
is important characteristic of the phase coexistence. Such a
quantity measures the asymmetry of the phases with respect to the
order parameter chosen. This asymmetry is related to the
particle-hole asymmetry \cite{crit_griffitswheeler/pra/1970} and is
sensitive to the interparticle correlations which governs the
expansion rate of the phases below the CP.

Let us make some general comments about the temperature behavior
of the rectilinear diameter. Obviously one can distinguish between
three characteristic regions: a) fluctuation region $T_{Gi}<T<T_c$
where $T_{Gi}$ is the Ginzburg temperature \cite{ll5}; b) normal states near
the triple point $T \gtrsim T_{tr}$; c) the intermediate one
$T_{tr} < T < T_c$. The temperature dependence of the rectilinear
diameter of the density is governed by the different physical
mechanisms in theses regions. Near the triple point the hard core
interactions in liquid phase play the dominant role so one can
expect that the behavior in this region should be determined by
the hard core interactions only. The near critical behavior is
governed by long range fluctuations. The specific temperature
behavior in the intermediate region depends on the balance of the
short range and long range correlations. Thus the rectilinear
diameter characterizes quantitatively to what extent the rate of
short range ordering in liquid phase overcomes the rate of
disorder in coexisting vapor.

\begin{figure}
  \centering
\subfigure[\,\,$Ar,Kr,Xe$]{\includegraphics[scale=0.55]{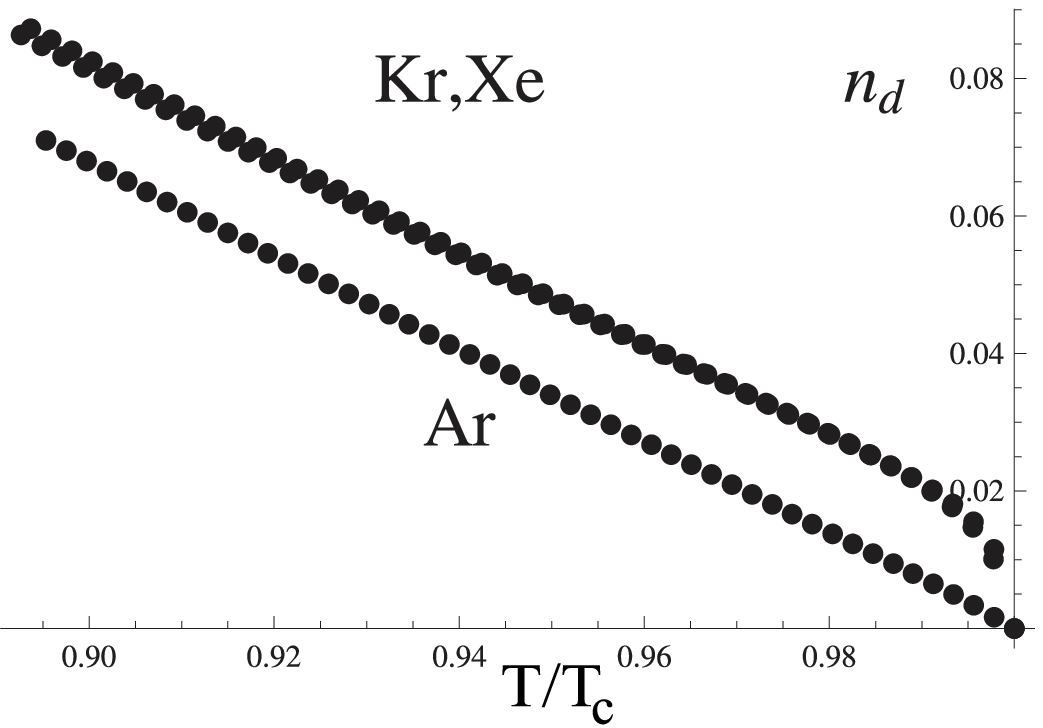}} \qquad
\subfigure[\,\,$He,H_2,Ne$]{\includegraphics[scale=0.55]{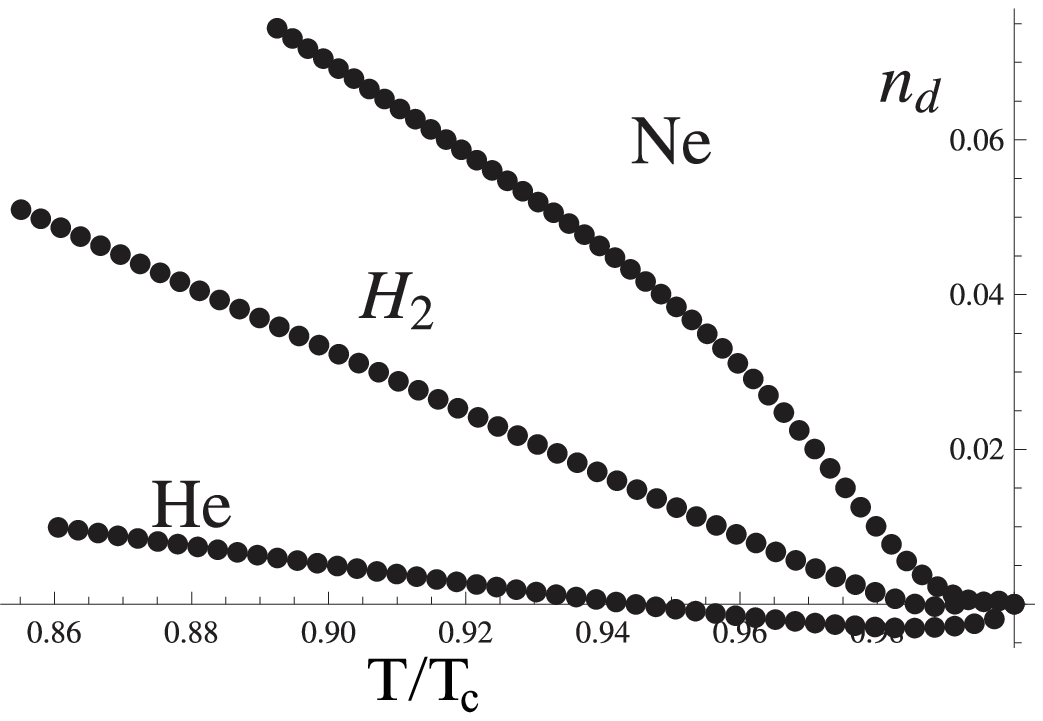}}
  \caption{The rectilinear diameter for the density $n_{d}$.}\label{densdiam}
\end{figure}

These general remarks are illustrated by the data for the
rectilinear diameter of the density (see Fig.~\ref{densdiam}). The
quantity $\rho_{d}$ is sensitive to the order in the local
configuration. In mono atomic liquids like noble ones the
monotonic change of this quantity takes place. This is because of
the stronger correlations in liquid phase. The role of the long
range correlations in the formation of non monotonic behavior of
$n_{d}$ in the near critical region is discussed below in
Section~\ref{sec_diam2}.

In the fluids with polyatomic molecules the volume available for
the thermal motion of the molecule modifies the rotational motion
in the effective ``cavity``. It leads to the fact that the
contribution of the inner degrees of freedom, in particular the
rotational ones, becomes dependent on the specific volume. As
a consequence the critical amplitudes should be different and this
is demonstrated by the behavior of the rectilinear diameter in near critical region (see
Fig.~\ref{densdiam}).\label{diamnote}
Note that the temperature behavior of the rectilinear diameter of
the density is the fine criterion for the validity of the EOS.

%%%%%%%%%%%%%%%%%%%%%%%%%%%%%%%%%%%%%%%%%%%%%%%%%%%%%%%%%%%%%%%%%%%%%%%%%%%%%%%%%%%%%%
%%%%%%%%%%%%%%%%%%%%%%%%%%%%%%%%%%%%%%%%%%%%%%%%%%%%%%%%%%%%%%%%%%%%%%%%%%%%%%%%%%%%%%

\begin{figure}
  \centering
\subfigure[\,\,$Ar,Kr,Xe$]{\includegraphics[scale=0.55]{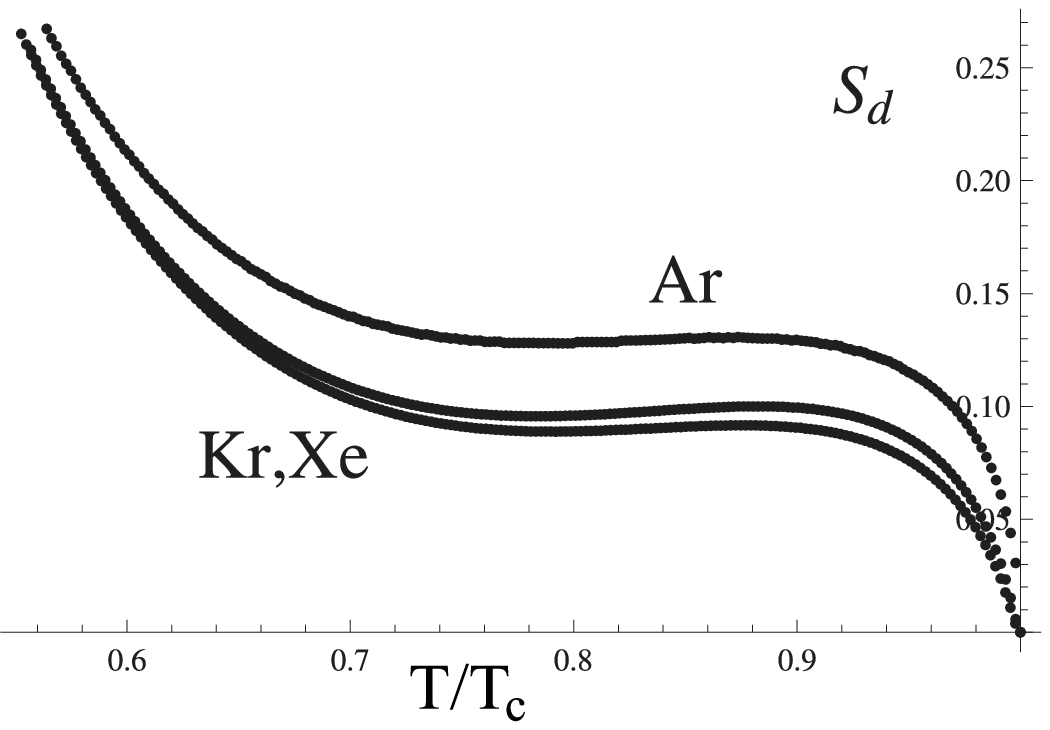}} \qquad
\subfigure[\,\,$He,H_2,Ne$]{\includegraphics[scale=0.55]{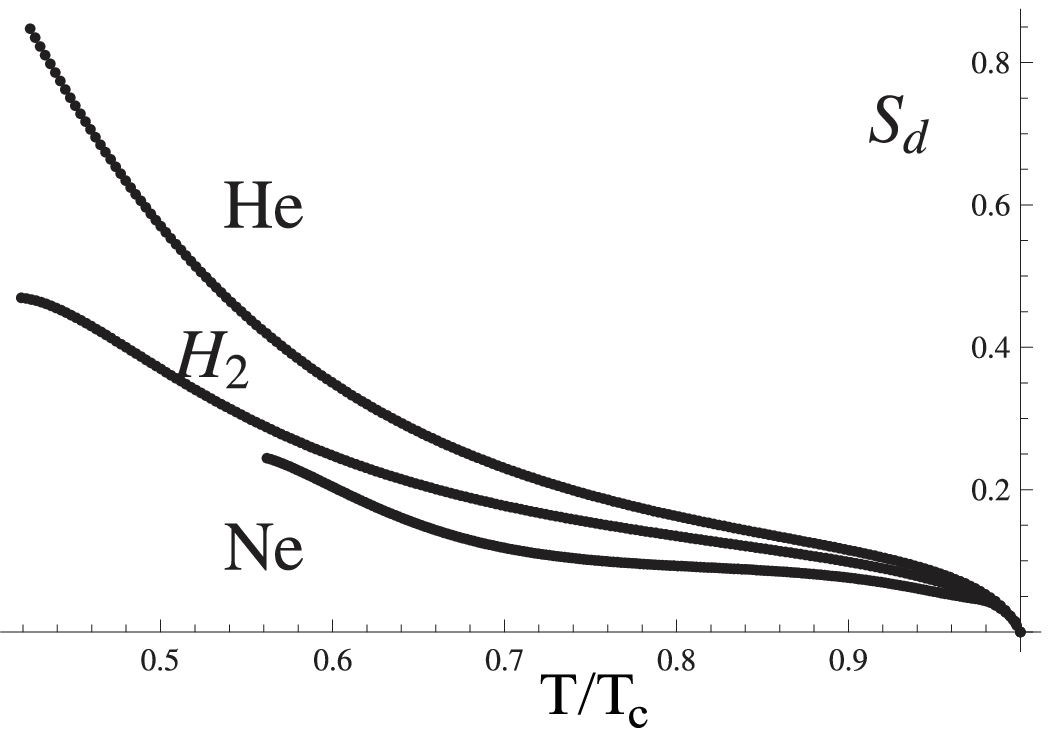}}
  \caption{The rectilinear diameter for the entropy $S_d$.}\label{entrdiam}
\end{figure}

The natural characteristic of the ordering process in the
coexistent phases is the entropy. Therefore the qualitative
difference in the ordering in coexisting phases can be also
characterized by the rectilinear diameter for the specific (per
particle) entropy:
\begin{equation}\label{diamentr}
  S_{d} = \f{S_{l}+S_{g}}{2} - S_c\,.
\end{equation}
The behavior of $S_d$, as well as $n_{d}$, for different
substances which obey the PCS should be the same.
For comparison on Figs.~\ref{entrdiam} we show the data $S_d$ for
noble fluids.
%%%%%%%%%%%%%%%%%%%%%%%%%%%%%%%%%%%%%%%%%%%%%%%%%%%%%%%%%%%%%%%%%%%%%%%%%%%%%%%%%%%%%%
%%%%%%%%%%%%%%%%%%%%%%%%%%%%%%%%%%%%%%%%%%%%%%%%%%%%%%%%%%%%%%%%%%%%%%%%%%%%%%%%%%%%%%
\begin{figure}
\centering
\subfigure[][The behavior of the residual entropy $S^{(int)}$ of the noble gases.]{
  \includegraphics[scale=0.6]{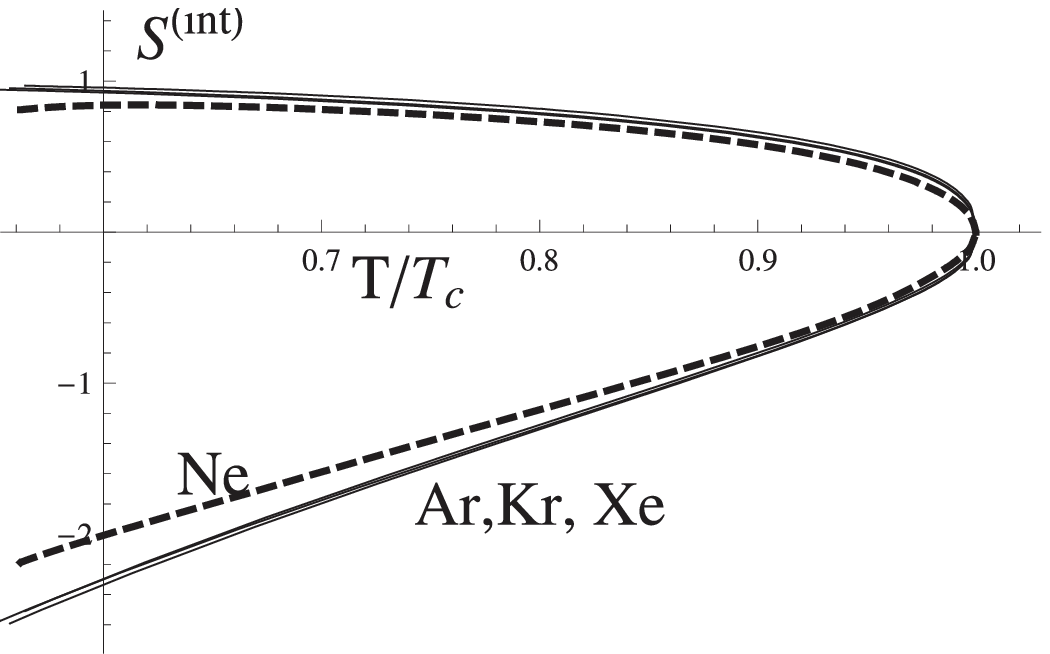}\label{fig_sintnoble}}\hspace{1cm}
%  \caption{The behavior of the residual entropy of the noble gases $Ne$ (doted) and $Ar,Kr,Xe$ %(bold).}
 \subfigure[][The behavior of the diameters for the total and residual entropy]{\includegraphics[scale=0.6]{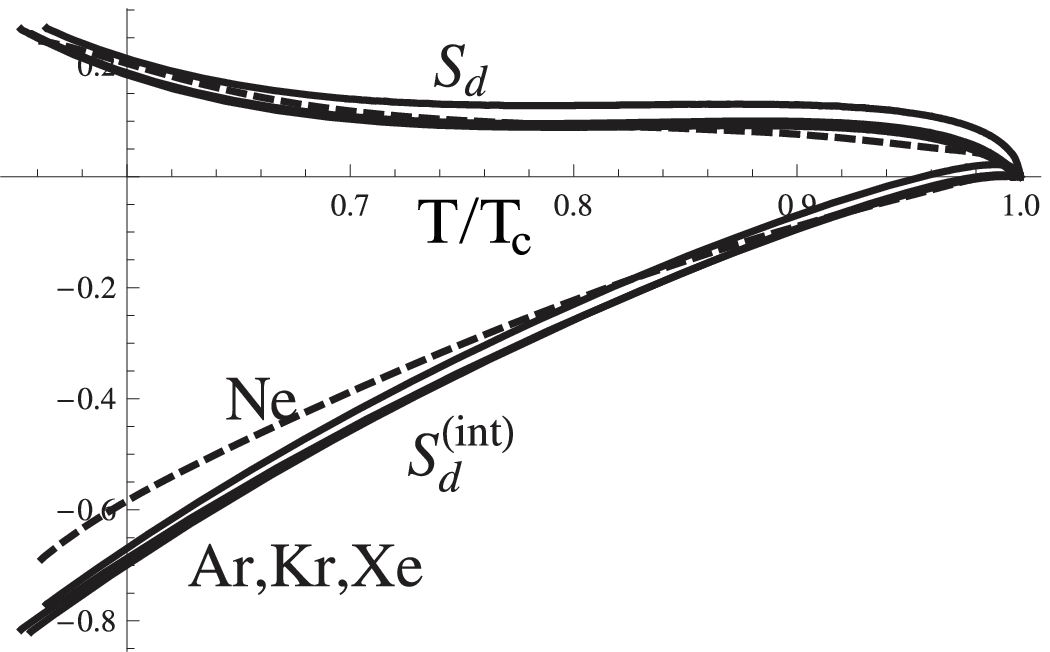}}
  \caption{The rectilinear diameter $S_{d}$ of the total entropy and the residual one $S_{d}^{(\text{int})}$ of the noble gases $Ne$ (doted) and $Ar,Kr,Xe$ (bold).}\label{snoblediam}
\end{figure}
Except the critical region the behavior of $S_d$ within the mean field approximation differs from the experimental curve by the shift of order $\Delta S$. The appearance of this difference is naturally explained by the the fluctuation shift of the critical temperature: \[\Delta S = S^{(mf)} - S_c  \approx c^{(mf)}_v\ln{\f{T^{(mf)}_{c}}{T_c}}\approx \f{T^{(mf)}-T_{c}}{T_c}\] where $c^{(mf)}_v$ is the dimensionless value of the mean field value of the specific heat, which for the heavy noble fluids takes the value $c^{(mf)}_v \approx 2$).
So
that the entropy diameter value at the CP  is less than that in the mean field approximation.

Let us state the relation between the behavior of diameters of the entropy $S_{d}$ and the density $n_d$.
We start from the basic thermodynamical representation of the entropy in dimensionless units (in units of the Boltzmann constant $k_{B}$):
\begin{equation}\label{entrop}
  S = S_c+ c_{v}\ln\f{T}{T_c} + f(n)-f(n_c)\,,
\end{equation}
where $S_c$ is the entropy at the CP, $c_v$ is the dimensionless specific heat, $f(n)$ is some function describing the density dependence of the entropy. From Eq.~\eqref{entrop} one can conclude that:
\begin{equation}\label{entropd}
  S_{d} = \f{c^{(l)}_{v}+c^{(g)}_v}{2}\ln\f{T}{T_c} +\f{f(n_l)+f(n_g)}{2} - f(n_c)\,.
\end{equation}
In the mean filed region below the CP we have:
\begin{equation}\label{binodens}
\f{\delta n_{l,g}}{n_c}= \pm\, b\sqrt{|\tau|}+a|\tau|+\ldots\,,\quad \tau  = \f{T-T_c}{T_c}<0
\end{equation}
where $\delta n = n-n_c$, and the indices $l,g$ stand for the coexisting liquid and gas phases correspondingly.
%and the specific heat is finite.
The searched relation between $S_d$ and $n_d$ near the CP is obtained by the expansion of Eq.\eqref{entropd} with respect to the density:
\begin{equation}\label{entropd1}
  S_{d} = n_c\left.\frac{\partial\,S}{\partial\, n}\right|_{c}\,n_d+\f{c^{(l)}_{v}+c^{(g)}_v}{2}\ln\f{T}{T_c} +
  \f{1}{4}\left(\left.\frac{\partial\,^2 S}{\partial\, n^2}\right|_{l}\delta n_l^2 + \left.\frac{\partial\,^2 S}{\partial\, n^2}\right|_{g}\delta n_g^2\,\right)+ \ldots\,.
\end{equation}
Further we use the thermodynamic relations \cite{ll5}:
\begin{align}
\left.\frac{\partial\, S}{\partial\, n}\right|_c =& -\f{1}{n^2_c}
\left.\frac{\partial\, p}{\partial\, T}\right|_c <0\,,\label{thermodynrelations1}\\
\left.\frac{\partial\,^2 S}{\partial\, n^2}\right|_c  =& \f{2}{n^3_c} \left.\frac{\partial\, p}{\partial\, T}\right|_c - \f{1}{n^2_c}  \left.
\frac{\partial\,^2 p }{\partial\, T\partial\, n}\right|_c
\label{thermodynrelations2}
\end{align}

Omitting the irrelevant terms near the critical point we get:
\begin{equation}\label{entropdfin}
  S_{d} =- \left[\ \f{1}{n_c}\left.\frac{\partial\, p}{\partial\, T}\right|_c\,\left(a-
  \,b^2\,\right)+ b^2\frac{\partial\,^2 p }{\partial\,n\,\partial\, T}
 + \f{c^{(l)}_{v}+c^{(g)}_v}{2}\right] |\tau |+ o(|\tau|) \,.
\end{equation}
The coefficient $b$ depends on the parameters of the EOS. In particular for the vdW EOS $a=\f{2}{5}, b=2$ and $\f{dS_d}{d\tilde{T}} = 0.6$, for the Bertlo EOS $a=\f{2}{5}, b=2\sqrt{2}$ with $\f{dS_{d}}{d\tilde{T}}=1.35$ in near critical mean-field region.
The slope of $S_d$ increases with the density asymmetry which is described by the coefficient $a$. For lighter noble fluids the slope of the entropy diameter is opposite to that for the heavy noble fluids. This is because of the fact that for light noble fluids the density asymmetry is weak (see Fig.~\ref{densdiam}) and the specific heat for the liquid phase \cite{pinesnozieres} is smaller in comparison with that for the heavy noble ones (see Fig.~\ref{diamentr}).

The difference  between the behaviors of $S_d$ for light and heave noble gases is
explained  by the dimerization effects (see Fig.~\ref{entrdiam}).
The calculation of the diameter for the total entropy $S_d$ as well as for its the residual part:
\[S^{(int)}_d = \f{S^{(int)}_l+S^{(int)}_g}{2}-S^{(int)}_c\,,\] where \[S^{(int)} = S-S^{(id)}\,,\qquad S^{(id)} = \f{3}{2}\,\ln\f{T}{T_c}-\ln\f{n}{n_c}\,,\] with the help of the vdW EOS, shows that the slope of the $S_d$ increases with account of the dimers (see remarks about Eq.~\eqref{entropdfin}). The comparison of the experimental data for $Ne$ and $Ar,Kr,Xe$ supports this conclusion (see Fig.~\ref{snoblediam})
%(see Fig.~\ref{sdiamvdwdimer})
 In order to clarify some qualitative details of the temperature behavior of $S_{d}$ we will discuss the dimerization of the heavy noble fluids in more details in the next Section. The behavior of $S_d$ in the fluctuation region we will discuss after this.

\section{The degree of the dimerization from the chemical equilibrium}\label{sec_dimer}
First let us make some general conclusions about the temperature  behavior of Eq.~\eqref{ration} along the binodal.
For the description of the dimerization let us introduce the quantity
\begin{equation}\label{rmd}
\mathfrak{R}_{dim} = \f{\tilde{n}}{\tilde{n}_1}\,,\quad \tilde{n} = \f{n}{n^{(c)}}\,,\quad \tilde{n}_1 = \f{n_1}{n^{(c)}_1}
\end{equation}
where $n$ is the total density (dimers and monomers) and $n_1$ is the number density for monomeric fluid.

For high enough density in liquid branch for the substances of
interest $\mathfrak{R}_{dim}$ will be close to unity far away from
the critical point. It is seems to be natural that with increasing
the density the dimerization becomes negligible.

At $T =  T_{c}$ according to the definition $\mathfrak{R}_{dim} =
1$. The monotonous change of $\mathfrak{R}_{dim}(T)$ between these
boundary values shows the quantitative similarity between
substances and thus the validity of the PCS. In contrast to this
the nonmonotonic behavior for $\mathfrak{R}_{dim}(T)$ along the
binodal signifies the qualitative difference in the process of
thermal expansion. According to what has been said above the
dimerization takes place at low enough densities near the critical
point. The formation of the dimers lead to the flattenning of the
binodal thus increasing the difference in the densities of the
coexisting phases. The increase of the density in liquid phase far
from the critical point makes the dimerization impossible both
from the energetic and configurational point of view. In such a
case one expects the nonmonotonous behavior of
$\mathfrak{R}_{dim}(T)$ along the binodal in liquid phase.

Here we calculate the dimerization degree for the heavy noble
fluids using the chemical equilibrium for the dimerization
process:
\[\text{monomer}+\text{monomer}\leftrightharpoons \text{dimer}\]
Let us introduce the degree of association $a$:
\begin{equation}\label{dimerizationdegree}
n_1 = (1-a)\,n_0\,,\quad n_2 = \f{a}{2}\,n_0\,.
\end{equation}
where $n_{2}$ is the number density of the dimers and $n_0 = n_1 +
2n_2$ is the total number density of atoms. The equation of
chemical equilibrium for $a$ is:
\begin{equation}\label{chemequil}
  \f{a}{(1-a)^2} =2 n_0\exp\left(\,\f{\,2\mu^{(ex)}_{1} - \mu^{(ex)}_{2}+E_{diss}}{T} \,\right)
\end{equation}
where $E_{diss}$ the dissociation  energy of the dimer (in calculation $E_{diss}=0.97\,T_c$.
The excessive chemical potentials are:
\[\mu^{(ex)}_{i} =\left.
\frac{\partial\, F^{(ex)}}{\partial\, n_i}\right|_{T}\,,\] where
$F^{(ex)}$ is the excessive free energy is the difference of the
total free energy $F$ and the ideal part $F^{(id)}$:
\begin{equation}\label{freemondim}
 F^{(ex)} = F - F^{(id)}\,.
\end{equation}
In its turn $F^{(ex)}$ can be decomposed
\[F^{(ex)} = F^{(hc)} + F^{(int)}\,,\]
where $F^{(hc)}$ is the hard core contribution and $F^{(int)}$ is
the contribution due to van der Waals interaction. Further we will
use the quantities reduced to the vdW values of the critical
parameters for monomeric fluid. The interaction term $F^{(int)}$
we take in standard binary form:
\begin{equation}\label{free12}
  F^{(int)} =-\f{9}{8}\left(\tilde{n}^2_1+\lambda \tilde{n}_1\,\tilde{n}_{2} +  \gamma \,\tilde{n}^2_{2}\right)\,.
\end{equation}
Assuming the similarity of the interparticle interactions (see
Eq.~\eqref{potential}) for monomers and dimers, within the vdW
approximation and $\lambda$ and $\gamma$ are determined as
following:
\[\lambda \approx \f{U_{12}}{8U_{11}}\left(1+\f{\sigma_2}{\sigma_1}\right)^3\,,\quad \gamma \approx
\f{U_{22}}{U_{11}}\left(\f{\sigma_2}{\sigma_1}\right)^3\,,\] where
$U_{11},U_{12},U_{22}$ are the interaction constants for
''monomer-monomer'', ''monomer-dimer'' and ''dimer-dimer''
correspondingly. In the order of magnitude $\lambda \gtrsim 1$ and
$\gamma \gtrsim 2$. The hard core contribution is taken in vdW
form with account of the mutual influence of excluded volume for
the monomers and the dimers:
\begin{equation}\label{hc}
  F^{(hc)} = -T \left[\,n_{1} \,\ln\left(\,1- \f{n_1}{3}  - \f{b_{1}\,n_2}{3}\,\right) + n_{2} \,\ln\left(\,1- \f{b_{2} n_2}{3}  -
  \f{b_{1}\,n_1}{3}\,\right)\,\right]\,.
\end{equation}
\begin{figure}
  \includegraphics[scale=0.7]{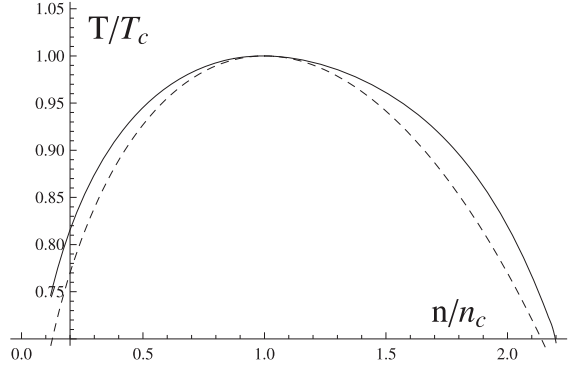}\\
  \caption{The binodals for vdW EOS (dashed) and for vdW with account of the dimerization
  (bold)}\label{binodal1}
\end{figure}
Here $b_{1}$ and $b_{2}$ are the specific volumes per dimer (in
units of the monomer volume) in environment of the monomers and
the dimers correspondingly.
%
%\begin{figure}
%  % Requires \usepackage{graphicx}
%  \includegraphics[scale=0.7]{r}\\
%  \caption{The corresponding ratio $\mathfrak{R}_{dim}$, the parameters $b_1= 1,\,b_d = 3,\,\gamma =2 ,\, \lambda
%  =1$}\label{rndim}
%\end{figure}
%
As the calculations show that the account of the dimerization
leads to the broadening of the binodal in $n-T$ coordinates (see
Fig.~\ref{binodal1}).

\begin{figure}
  \includegraphics[scale=0.7]{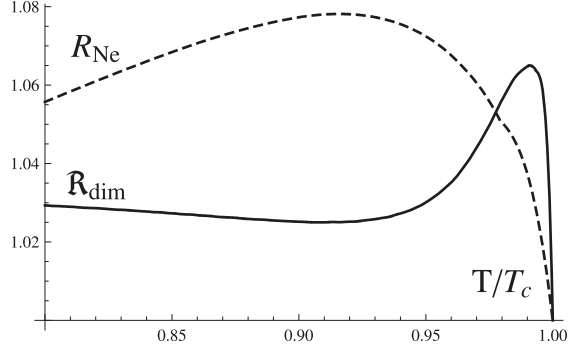}\\
  \caption{The comparison of the theoretical (dashed) result for
  $\mathfrak{R}_{dim}$ and the experimental $R(Xe,Ne)$ (bold) data (the parameters $b_1= 1,\,b_2 = 3,\,\gamma =2 ,\, \lambda
  =1$).}\label{rcompar}
\end{figure}

The ratio $\mathfrak{R}$ is shown in Fig.~\ref{rcompar}. The
comparison with the experimental data of $R_{Ne}$ for the pair of substances $Xe$ and $Ne$ as the vdW fluids with dimers and without them correspondingly. It  shows that the mean field
approach is adequate at least qualitatively (see
Fig.~\ref{rcompar}).

The nonmonotonic behavior of $R_{Ne}$ as it follows from
Fig.~\ref{nobleref} takes place in close vicinity of the critical
point. The formation of the dimers becomes impossible if the
specific volume becomes less that $2v_0$ (in reduced coordinates
$\tilde{n}\approx 1.5$). Simple estimate shows that it occurs
exactly at the temperature $T\approx 0.95 \div 0.97\, T_c$. In the
proposed mean field description $\mathfrak{R}_{dim}$ changes in
broader region, $\tilde{n}\approx 2$ (see Fig.~\ref{binodal1}).
Such a discrepancy can be attributed to the inadequate account of
the hard core effects which mainly governs the destruction of the
dimers when the density increases. The use of more appropriate
approximation for the hard sphere contribution, e.g.
Carnahan-Starling, decreases the value of $\mathfrak{R}_{dim}$
beyond the fluctuation region.
The behavior of $S_{d}$ in the mean field region, $T/T_c < 0.9$,
depends on the short range correlations. Obviously the difference
in the character of such short range correlations is responsible
for the different behavior of the rectilinear diameter of the
entropy $S_{d}$ which is clearly shown on Fig.~\ref{entrdiam} \,for
the heavy ($Ar,Kr,Xe$) and light noble fluids. As has been noted
above the existence of the dimers is not possible in this region.
But the stronger short range correlations, as compared to those in
lighter ones, should be expected in the heavy noble gases as the
remnants of the near critical point dimerization.

\section{The rectilinear diameter anomaly}\label{sec_diam2}
The asymptotic behavior of the rectilinear diameter of the density in the fluctuational region is as following:
\begin{equation}\label{densdiamcrit}
  n_{d} = \f{n_{l}+n_{g}}{2n_c} - 1 = D_{2\beta}\,|\tau|^{2\beta} +
  D_{1-\alpha}\,|\tau|^{1-\alpha} + D_1 |\tau| + o(\tau)\,,\quad D_1<0\,.
\end{equation}
It contains $|\tau|^{2\beta}$-term \cite{crit_diamexp/molphys/1986,crit_can_kul/cmphukr/1997,crit_fishmixdiam1/pre/2003}. The
ground for introduction of such a term as the sequence of the
choice of the order parameter predicted firstly within the
canonical formalism approach
\cite{crit_can_kul/cmphukr/1997,crit_coulomb_critelectrolyte_kulimalo/pre/1999}.
Note that according to \cite{crit_can_diamsing_kulimalo/physa/2008} the ratio
of the leading singularity amplitudes for the temperature
dependence of the rectilinear diameter is ``almost`` universal:
\begin{equation}\label{ratioa}
  \f{D_{1-\alpha}}{D_{2\beta}} \simeq
  a^{1-\alpha -2\beta}\,\f{l_s(0)}{g^2_s(0)}<0\,.
\end{equation}
where $a$ is the nonuniversal factor. The quantities $f_s$ and $g_s$ are universal functions:
\[f_{s}(x) = f_{s}(-x)\,,\quad g_{s}(x) = \f{d\,f_{s}(x)}{dx}\,.\]
They are determined by the asymptotic thermodynamic potential
\[\Phi = |A_2|^{2-\alpha}\,f_s\left(\,\f{A_1}{A^{\beta + \gamma}_2}\,\right)+\ldots\]
in isomorphic variables $A_1$ - conjugated field-like variable and $A_2$ - temperature-like variable.

The temperature behavior of the rectilinear diameter strongly
depends on the interplay between the values of $D_{i}$. Basing on the experimental data shown in Fig.~\ref{densdiam} it seems
natural to assume that $D_{1} > 0 $. From the physical
interpretation of the $n_{d}$ it is clear that the repulsive
terms in pressure decrease the absolute value of $D_1$. Since the
$2\beta$ singularity is the strongest one the non monotonous
behavior of the density rectilinear diameter takes place if
$D_{2\beta}<0$. In order this behavior manifests clearly there
should be $D_{1-\alpha}< |D_{2\beta}| $ which according to
Eq.~\eqref{ratioa} means that $a<1$. The parameter $a$ depends on
the choice of the initial order parameter. This explains the fact
that $2\beta$ singularity amplitude varies greatly depending on
the order parameter chosen. This usually is interpreted as
the division of the order parameters into more and less symmetrical ones, e.g. like
density $\rho$ and volume $v$ \cite{henselknuth/prl/1985}.

Even if $D_{2\beta}>0$ the $2\beta $ singularity can be masked by
the interplay of linear and ``$1-\alpha$`` terms $D_{1}\approx
|D_{1-\alpha}|$ if $D_{2\beta}/|D_{1-\alpha}|\ll 1$ ($a<1$, see
Eq.~\eqref{ratioa}). In particular from the results of
\cite{crit_fisherdiam/chemphyslet/2005} it follows that
$\f{D_{2\beta}}{D_{1-\alpha}} \simeq -0.1$ with $D_{2\beta}>0$.
Note that in \cite{crit_fisherdiam/chemphyslet/2005} the data for the
liquids with different degree of the binodal were analyzed, namely
$SF_6$ as typical molecular liquid and alkali metals $Cs, Rb$.

If $D_{2\beta}>0$ and in addition $(1-\alpha)|D_{1-\alpha
}|\gtrsim D_{1}$ then essentially nonmonotonous behavior for the
rectilinear diameter takes place. There is minimum at
\[\tau_{min}\approx \f{1}{\alpha}\,\ln\f{(1-\alpha)
D_{1-\alpha}}{D_{1}}\] and the maximum \[\tau_{max}\approx
\left(\,\f{2\beta}{1-\alpha }\f{ D_{2\beta}}{D_{1-\alpha}}
\,\right)^{\f{1}{1-2\beta -\alpha}}\,.\] The location of this
maximum is very close to $T_c$ ( $\left|D_{2\beta
}/D_{1-\alpha}\right|<1$) and can be measured only if
$(1-\alpha)|D_{1-\alpha }|\approx D_{1}$. The clear example of
such situation is given by rectilinear diameter for the binodal of
$Hg$ \cite{diamliqmetals_hensel/jphys/1996}
%(see Fig.~\ref{hgdiam})
.
The data used for light nobel fluids here does
not allow to distinguish between this case and the situation with
$D_{2\beta}<0$.
%\begin{figure}
%  % Requires \usepackage{graphicx}
%  \includegraphics[scale=0.8]{hg_densdiam}\\
%  \caption{The rectilinear diameter for $Hg$ and $Rb$ (picture taken from
%  \cite{diamliqmetals_hensel/jphys/1996}).}\label{hgdiam}
%\end{figure}

The values of the amplitudes $D_{2\beta}$ and $D_{1-\alpha}$ are
the same for the substances which belong to the same class of
corresponding states (eg. $Kr$ and $Xe$). As is shown in
\cite{crit_can_diamsing_kulimalo/physa/2008} $D_{2\beta}$ and $D_{1-\alpha}$
have opposite signs. The coefficient $D_{2\beta}$ can be either
positive or negative depending on the details of the
intermolecular interactions. Taking into account the results of
the previous section we can conclude that for molecular liquids as
well as in liquid metals \cite{diamliqmetals_hensel/jphys/1996}
where the clusterization takes place in near critical region the
coefficient $D_{2\beta}$ is positive. The same is true for the
entropy diameter and any other order parameter for which the
Landau-Ginsburg functional contains the odd power terms.

The situation with the rectilinear diameter for lighter noble
fluids suggests two possibilities. The first one is that the
amplitude $ |D_{1-\alpha }| > (\gg)D_{2\beta}>0$ but
$(1-\alpha)|D_{1-\alpha }|\gtrsim D_{1}$ and the binodal is quite
symmetrical. The second is that $a<1$ so that $D_{1-\alpha}(\ll) <
- D_{2\beta}$. The data for $Ne$ also conform with such a
proposition. Note that helium has ``almost`` symmetrical binodal
($|D_1| \ll 1$) in comparison with other noble liquids. Obviously
this can be attributed to the quantum corrections to the equation
of state which gives the additional repulsive contribution to the
pressure. Thus one can expect that in $He$ the coefficient
$D_{2\beta}$. From here we can make the following conclusion
concerning the sign of the amplitude $D_{2\beta}$. The
dimerization influences the character of the short range
correlations. If the dimerization is negligible then one can
expect that $D_{2\beta}<0$. For the systems where the dimerization
takes place $D_{2\beta}>0$. This leads to the qualitative
difference in the behavior of the rectilinear diameter $n_{d}$
observed.

Here we perform the analysis of the critical asymptotic of the rectilinear diameter for the entropy
\begin{equation}\label{entropdiamcrit}
  S_{d} = S_{2\beta}\,|\tau|^{2\beta} +
  S_{1-\alpha}\,|\tau|^{1-\alpha} + S_1 |\tau| + o(\tau)\,.
\end{equation}
The first contribution is caused by the corresponding term
of the density diameter and the quadratic density contribution in Eq.~\eqref{entropd1}.
Note the term $\propto \tau^{1-\alpha}$ is generated by the density
diameter and the specific heat terms according to Eq.~\eqref{entropd1}. Note, that as follows from the representation \eqref{entrop} $S_{1-\alpha} < 0 $. In contrast to the coefficients $D_{1-\alpha}$ which is generated by the asymmetry of the Hamiltonian, the coefficient $S_{1-\alpha}$ does not vanish in symmetrical case, e.g. in the Ising model, and correspondingly:
\begin{equation}\label{sstruct}
 S_{1-\alpha} = S^{(sym)}_{1-\alpha} + S^{(asym)}_{1-\alpha}\,.
\end{equation}
It is obvious that the symmetrical part $S^{(sym)}_{1-\alpha}$ is generated by the heat capacity singularity.

As follows from Eq.~\eqref{entropd1}, $S_{1-\alpha} < 0 $. In
order to distinguish between the $\tau^{2\beta}$ and
$\tau^{1-\alpha}$ terms for rectilinear diameter we subtracted the analytic background terms outside the critical region. The coefficients $s_k$ of the polynomial \[S^{reg}_d = \suml_{k=1}^{5} s_k \left(\,T_c/T -1\,\right)^k\] have been determined with the help of the experimental data in the region $T/T_c< 0.75$.
%taking into account the data beyond the region $|\tau| > t_{*}$. Such a subtracting allow to fix approximately the value $t_{*}$ because the linear analytic term is important even in fluctuational region. We fit the data in the region $\tau>t_{*}$ with the polynomial expansion in powers of $\tau/1+\tau)$ up to $5$-th order so that no significant oscillating deviation between the regular expansion and the data occurs. In the framework of such procedure the value $t_{*}=0.25$ appeared to be optimal so that no significant deviations in the region $\tau > t_{*}$ between regular expression and the data were detected.
The results of calculation of the ratio $S_{2\beta}/S_{1-\alpha}$ are placed in Table~\ref{tabent} and they confirm the conclusion of the canonical formalism about its universal character. Small variations in this ratio in dependence on the specific fluid are explained by
the difference between canonical temperature variable $A_{2}$ and the reduced temperature $\tau $. Besides we neglected the symmetrical in $S^{(sym)}_{1-\alpha}$ since it is proportional to the fluctuation correction of the heat capacity. The results allow to conclude that $S^{(sym)}_{1-\alpha}$ is smaller that the term generated by the asymmetry of the Hamiltonian.
\begin{table}
  \centering
\begin{tabular}{|c|c|c|c|c|c|c|c|c|c|c|c|c|}
\hline
   & $H_2$ & $Ne$ & $Ar$ & $Kr$ & $Xe$ \\
   \hline
     $\f{S_{2\beta}}{S_{1-\alpha}}$ &-0.28& -0.30& -0.31& -0.34& -0.33\\ \hline
\end{tabular}
  \caption{The ratio $\f{S_{2\beta}}{S_{1-\alpha}}$ for the noble gases.}\label{tabent}
\end{table}
\section{Discussion}
In this work it is stressed that the leading role in the understanding of the difference
and similarity in the thermodynamic behavior of fluids belongs to the PCS.
The main result of the present work is the clear demonstration of that fact,
that noble gases do not rigorously follow the PCS principle of corresponding
states \cite{book_prigozhisolut}.

As was shown from the analysis of the density and the entropy
the deviations from PCS are observed 1) for liquid phase near
the critical point and 2) and for gaseous phase in the whole region of their
existence. The latter is especially important circumstance, which directly indicates the main cause of such deviations. It is the dimerization
processes in noble gases. The most essentially the dimerization manifests itself in gaseous phase far
away from the critical point, where the doubled fraction volume per molecule $2\upsilon $ is noticeably more than the own volume for a dimer. The degree of the dimerization here reaches $10\%$. In liquid phase due to the geometrical restriction for the existence of the dimers the temperature interval where the dimerization effects in nobel fluids take place is rather narrow due to the closeness of the dissociation temperature $T_{diss}$ to the critical one $T_c$ ($T_{diss}\approx 0.96\, T_c$). Nevertheless they are quite noticeable in the temperature behavior of the entropy and the specific heat. The analysis of the experimental data with taking into account of the dimerization effect for the liquid phase in near critical region give the estimate $\lesssim 5\%$ for the degree of dimerization.
Note that these effects can not be adequately described with the standard two
parametric vdW-like EOS.

The density and the entropy diameters are the fine characteristics of the EOS. From our analysis
it follows that widely used vdW and Bertlo EOS are not able to describe the
formation of the plato in the temperature dependence of $S_d$ for the heavy noble fluids.
At the same time the density binodal is reproduced quite correctly.

The analogous situation is characteristic for liquids with nonspherical molecules, first of all, for benzene and water.
In these cases the behavior of the entropy diameter is essentially nonmonotonic. This circumstance for water is expected since the association processes in it are considerably more expressed than in noble gases. In \cite{water_dimer_nato/2007/us} it was shown that the degree of the dimerization near the critical point is close to unit ($a\approx 0.9$). At removing from the critical point the associates in liquid water becomes more complicated.
So their influence on the fine details of density and the entropy of water is considerably more expressed. Strong correlations in liquid benzene lead also to considerable
changes in the behavior of $S_d$. The specific structure manifest itself also in
kinetic properties such as viscosity.

As shown in the work, the dimerization radically influences the behavior
of the density rectilinear diameter. From the analysis of the last for $He$, $H_{2}
$ and $Ne$ it follows that $D_{2\beta}  $ is negative or takes the small
positive value, essentially less than $D_{1 - \alpha}  $. Note, that the
using of the experimental data from \cite{crit_fisherdiam/chemphyslet/2005}
allows to get the following estimate
for their ratio: $D_{2\beta}/D_{1 - \alpha}  = - 0.1$.

It is necessary to
emphasize that in accordance with the canonical formalism reasons the signs
of $D_{2\beta}$ and $D_{1 - \alpha}$ should be opposite. This very
important conclusion could not be got with the help of theory developed in
\cite{crit_fishmixdiam1/pre/2003,crit_fishmixdiam2/pre/2003}.
This question will be discussed in separate paper.

One of the author (V.L.K.) acknowledge the partial financial
support of the INTAS (Project No. 06-1000012-8707).

%\bibliography{criticality,water,hbond,liqmetals,dimers,books_my}

\end{document}